\documentclass[aps,prl,twocolumn,groupedaddress,showpacs,nofootinbib]{revtex4}
\usepackage{graphicx}
\usepackage{times}
\def\beq{\begin{equation}}
\def\eeq{\end{equation}}
\def\bey{\begin{eqnarray}}
\def\eey{\end{eqnarray}}

\def\lsim{\mathrel{\raise.3ex\hbox{$<$\kern-.75em\lower1ex\hbox{$\sim$}}}}
\def\gsim{\mathrel{\raise.3ex\hbox{$>$\kern-.75em\lower1ex\hbox{$\sim$}}}}

\begin{document}

\title{High Energy Neutrinos As A Test of Leptophilic Dark Matter}
\author{Douglas Spolyar$^{1,2}$, Matthew Buckley$^3$, Katherine Freese$^4$, Dan Hooper$^{1,5}$, and Hitoshi Murayama$^{6,7,8}$}
\affiliation{$^1$Center for Particle Astrophysics, Fermi National Accelerator Laboratory, Batavia, IL 60510 USA}
\affiliation{$^2$University of California, Santa Cruz, Physics Department,
Santa Cruz, CA 95064 USA}
\affiliation{$^3$Department of Physics, California Institute of Technology, Pasadena, CA 91125, USA}
\affiliation{$^4$Michigan Center for Theoretical Physics,
 Physics Dept.,  University of Michigan, Ann Arbor, MI 48109, USA}
\affiliation{$^5$Astronomy and Astrophysics Department, University of Chicago, Chicago, IL 60637 USA}
\affiliation{$^6$Department of Physics, University of California, Berkeley, CA 94720, USA}     
\affiliation{$^7$Theoretical Physics Group, LBNL, Berkeley, CA 94720, USA}
\affiliation{$^8$IPMU, University of Tokyo, 5-1-5 Kashiwa-no-ha, Kashiwa,
                Japan 277-8568}
                \date{\today}
\begin{abstract}

Recently, observations by PAMELA, the Fermi Gamma Ray Space Telescope, and other cosmic ray experiments have generated a great deal of interest in dark matter particles which annihilate at a high rate to leptons. In this letter, we explore the possibility of using large volume neutrino telescopes, such as IceCube, to constrain such models.  We find that IceCube (in conjunction with the planned low threshold extension, DeepCore) should be capable of detecting neutrino-induced showers from dark matter annihilations taking place in the inner Milky Way in a wide variety of models capable of producing the excess reported by the Fermi Gamma-Ray Space Telescope and PAMELA. If Dark Matter annihilations are responsible for the signals observed by both PAMELA and FGST, then IceCube/DeepCore should detect or exclude the corresponding neutrino signal from the inner Milky Way with a few years of observation. If only the PAMELA signal is generated by dark matter annihilations, IceCube/DeepCore will be able to place stringent constraints on the fraction of dark matter annihilations that proceed to muons, taus, or neutrinos.
\end{abstract}
\pacs{95.35.+d;95.30.Cq,98.52.Wz,95.55.Ka
\hspace{0.5cm} FERMILAB-PUB-09-269-A
\hspace{0.5cm} CALT-68-2740}
\maketitle

Weakly Interacting Massive Particles (WIMPs) are among the best motivated classes of candidates for the dark matter (DM) of our universe (for reviews, see Refs.~\cite{Jungman:1995df,Lewin:1995rx,Primack:1988zm,Bertone:2004pz}).
The search for these particles is one of the primary missions of the Large Hadron
 Collider at CERN.  Stable particles with weak-scale interactions and masses are naturally predicted to annihilate among themselves (or with their antiparticles) at a rate in the early universe that leads to a thermal abundance similar to the observed density of DM. This same annihilation process is also expected to be taking place in the present universe, potentially providing an opportunity for DM's indirect detection.  The products of WIMP annihilations consist of a combination of electrons, positrons, protons, antiprotons, photons, and neutrinos, each of which may be potentially detected in existing or planned experiments. Some of the most studied strategies for the indirect detection of DM include searches for neutrinos from the Sun~\cite{Srednicki:1986vj} or Earth~\cite{Freese:1985qw}, searches for gamma rays from the Galactic Center~\cite{gs} or dwarf 
 spheroidal galaxies~\cite{dwarf}, and charged cosmic rays from annihilations throughout the halo of the Milky Way~\cite{charged}.

  \begin{figure}[t]
\includegraphics[width=\columnwidth]{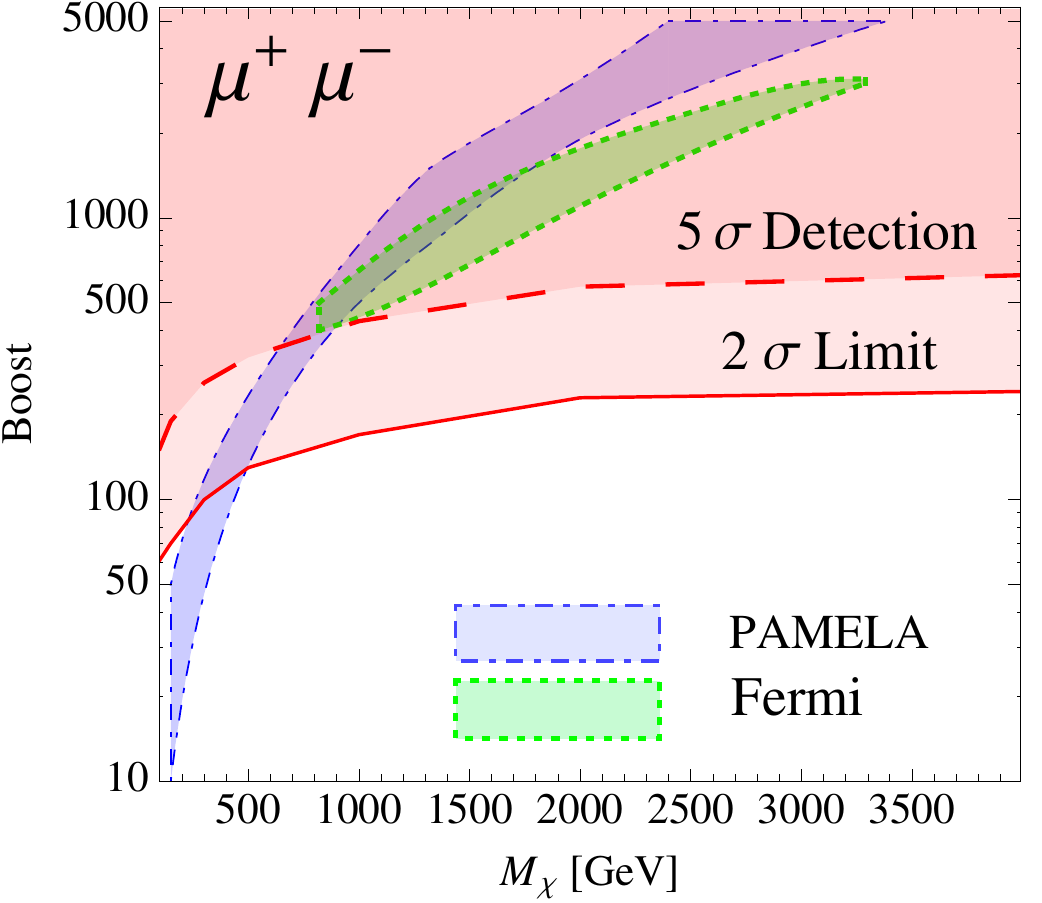}
\caption{Reach of ICECUBE/DeepCore to neutrinos from DM annihilation to $\mu^+ \mu^-$.  Shown are
the 5$\sigma$ (dashed) detection and the 2$\sigma$ limit (solid) on the boost factor as a function of WIMP mass after 5 years of operation.  Also shown are
the $2\sigma$ contours in the boost factor $B$ as function of DM mass for Fermi (dotted) and PAMELA (dot-dashed) inferred from~\cite{Bergstrom:2009fa} 
for $(\chi \chi\rightarrow \mu^+\mu^-)$.   }
\label{figure}
\end{figure}
  
Over the past several years, there have been a number of experimental signals which have been interpreted as possible indications of DM~\cite{Adriani:2008zr,Chang:2008zzr,Torii:2008xu,Collaboration:2009zk,Barwick:1997ig,ams,dama,integral,deboer,Elsaesser:2004ap,haze}. Confirmation that any of these observations are actually due to DM, rather than being a mere experimental artifact or astrophysical background, would likely require more than one experiment to provide complementary information.  In this letter, we consider the anomalous features in the spectrum of cosmic ray positrons and electrons reported by PAMELA~\cite{Adriani:2008zr}, ATIC~\cite{Chang:2008zzr}, PPB-BETS~\cite{Torii:2008xu}, and very recently by the Fermi Gamma Ray Space Telescope (FGST)~\cite{Collaboration:2009zk} (as well as in earlier indications from HEAT~\cite{Barwick:1997ig} and AMS-01~\cite{ams}). These observations have led to a great deal of speculation that DM annihilations~\cite{annihilation,annbf} or decays~\cite{decay} may be responsible. However, any explanation of these positron/electron signals in terms of DM annihilation requires somewhat nonstandard WIMP properties.  In particular, the local halo density of DM within the vicinity of the Solar System is insufficient to produce these observations unless the annihilation cross section is considerably larger than that typically expected for a thermal relic, or the annihilation rate is otherwise supplemented by a large boost factor $\sim 10^1 - 10^4$. Such a boost factor could plausibly arise due to particle physics such as a Sommerfeld enhancement~\cite{Sommerfeld}, in which the presence of an attractive potential between WIMPs leads to a low-velocity annihilation cross section that is enhanced relative to the value in the early universe. Alternatively, such an enhancement could arise due to astrophysics; for example, due to substructures in the DM distribution.  Furthermore, to explain the spectral shape reported by PAMELA and/or FGST, as
well as to avoid overproducing antiprotons (in excess of what is observed),
the DM annihilations must proceed largely to leptons. This is in contrast to standard 
neutralino DM, which typically annihilates to heavy quarks and gauge bosons. 

To confirm that the PAMELA and/or FGST signals arise from the annihilations of a leptophilic DM particle, one would hope to observe multiple species of annihilation products. In particular, in addition to positrons/electrons, one could observe gamma-rays and high energy neutrinos.  In this letter, we discuss neutrinos in IceCube as a possible test for DM annihilating to leptons in the halo of the Milky Way. While neutrinos from WIMP annihilations in the halo of our galaxy have been studied previously~\cite{Yuksel:2007ac}, there are new implications for this detection channel in light of the high annihilation rate and preference for leptonic modes required to explain the PAMELA and other anomalous cosmic ray signals. Furthermore, with the planned addition of DeepCore within the IceCube detector, it will become possible to observe neutrino-induced showers with only tens of GeV energy. Without DeepCore, using IceCube to study the inner Milky Way is difficult because of the large background of downgoing atmospheric muons. With the addition of DeepCore, the IceCube detector itself can be used as a veto for muon backgrounds, making possible the identification of neutrino-induced showers from the DM annihilations in the inner Milky Way.



\begin{table*}
\begin{tabular}{|c|c|c|c|c|c|c|}
\hline

~$m_\chi$&~ Bin Size & ~5$\sigma$~Detection~ & ~2$\sigma$~Limit~ &~5$\sigma$~Detection &~ 2$\sigma$~Limit & Normalization Required by PAMELA+FGST\\

(GeV)&(GeV) &~($\mu^+\mu^-$)~&~($\mu^+\mu^-$)~& ~($\mu^+\mu^- +\nu_\mu\bar\nu_\mu$)~&($\mu^+\mu^- +\nu_\mu\bar\nu_\mu$)~& ($\mu^+ \mu^-$ case)\\ \hline\hline
2000& 600-3000 & $B\geq570$&$B\leq230 $& $B\geq350$&$B\leq140$&$B\approx$ 1700\\ \hline
1000& 1500-300 & $B\geq 430$& $B\leq170$& $B\geq270$&$B\leq110$&$B\approx$ 450-700\\ \hline
\end{tabular}

\caption{The capability of IceCube/DeepCore to detect neutrino-induced showers in a scenario in which the PAMELA and FGST signals are both the result of DM annihilations. To produce these observed
signals, the DM particle must be quite heavy, and annihilate preferentially to muons (as opposed to electrons or taus). We show results for DM which annihilates (i) entirely to muons and (ii) to an equal number of muons and muon neutrinos. The limits on the boost factor and the discovery prospects given correspond to five years of observation. We have used an angular window corresponding to half of the sky (2$\pi$ sr) and an effective volume for DeepCore of 0.04 km$^3$ (the remaining volume of IceCube is used only as a muon veto). Comparing the limits obtained to the boost factors required to normalize to the PAMELA and FGST signals (given in the last column and inferred from Ref.~\cite{Bergstrom:2009fa}), we find that IceCube should be capable of testing any DM scenario that is responsible for these observations of the cosmic ray electron and positron spectra. 
 \label{tab:capnumu}}
\end{table*}


\begin{table*}
\begin{tabular}{|c|c|c|c|c|c|c|c|c|}
\hline

~$m_\chi$&~ Bin Size & ~5$\sigma$~Detection~ & ~2$\sigma$~Limit~ & ~5$\sigma$~Detection~ & ~2$\sigma$~Limit~ &~5$\sigma$~Detection &~ 2$\sigma$~Limit & Norm. Required by PAMELA\\

(GeV)&(GeV) &~($\mu^+\mu^-$)~&~($\mu^+\mu^-$)~&~($\tau^+\tau^-$)~&~($\tau^+\tau^-$)~& ~($\mu^+\mu^- +\nu_\mu\bar\nu_\mu$)~&($\mu^+\mu^- +\nu_\mu\bar\nu_\mu$)~& \\ \hline\hline
500 & 150-800  & $B\geq 320$ & $B\leq 130$ & $B\geq 480$ & $B\leq 190$& $B\geq 200$&$B\leq80$&  \begin{tabular}{c} $B \approx$ 120-800 ($\mu^+ \mu^-$) \\ $B \approx $ 200-500 ($\tau^+ \tau^-$)\end{tabular} \\ \hline

300 & 100-500 & $B\geq 260$ & $B\leq 100$ & $B\geq 370$ & $B\leq 150$& $B\geq159$&$B\leq60$& \begin{tabular}{c} $B \approx$ 40-180 ($\mu^+ \mu^-$) \\ $B\approx$ 70-160 ($\tau^+ \tau^-$) \end{tabular} \\ \hline

150 & 50-250 & $B\geq190$ &$B\leq70$ & $B\geq270$ &$B\leq110$ &$B\geq110$&$B\leq40$& \begin{tabular}{c} $B \approx$ 10-50 ($\mu^+ \mu^-$) \\ NA ($\tau^+ \tau^-$) \end{tabular}\\ \hline
\end{tabular}
\caption{The same as in Table I, but only assuming that the PAMELA signal is produced by DM annihilations (and not necessarily FGST).  An additional annihilation channel to taus
is considered as well. For this case of explaining PAMELA only, IceCube/DeepCore should be able to 
exclude the possibility that WIMPs heavier than approximately 500 GeV are responsible for the PAMELA
 excess (to estimate the values in the last column, we used the results of Ref.~\cite{annbf}.\label{tab:capnumu}}
\end{table*}

The flux of DM annihilation products from the direction of the Galactic Center is given by 
\begin{equation}
\frac{d\Phi (\Delta \Omega,E)}{dE} = \frac{B}{8\pi} \frac{\langle \sigma v\rangle}{m_\chi^2} \bar{J}(\Delta \Omega)\Delta \Omega  \sum_i f_i  \frac{dN^i_{\nu}}{dE_{\nu}}, \label{eq:dPhidEdef}
\end{equation}
where $f_i$ is the branching ratio to a given species, $dN^i_{\nu}/dE_{\nu}$ is the differential neutrino spectrum per annihilating WIMP, $B$ is the boost factor, $m_\chi$ is the DM mass, and $\langle \sigma v\rangle$ the annihilation cross section.  The DM distribution integrated over the line-of-sight over a solid angle, $\Delta \Omega$, is given by
 \begin{equation}
 \label{j}
 J = \int_{l.o.s.} \rho_\chi^2(s) ds \hspace{1em};\hspace{1em}
 \bar{J}(\Delta \Omega) = 
 {1 \over \Delta \Omega} 
 \int_{\Delta \Omega} PSF \star J d\Omega \label{eq:Jdef}
 \end{equation}
 where PSF is the point spread function of the instrument. Throughout our study, we will assume $\langle \sigma v\rangle = 3 \times 
10^{-26}~\mbox{cm$^3$/s}$ (the standard estimate for the DM annihilation cross section for a WIMP that is thermally produced in the early universe). 

The spectrum of neutrinos resulting from a WIMP annihilation depends on the mass of the WIMP and on the dominant annihilation modes. WIMPs annihilating to muons produce muon and electron neutrinos in their decays.  Annihilations to taus produce neutrinos through a varity of decay processes, including $\tau \rightarrow \mu \nu \nu$, $e \nu \nu$, as well as from the hadronic decays $\tau \rightarrow \pi \nu$, $K \nu$, $\pi \pi \nu$, and $\pi \pi \pi \nu$~\cite{Jungman:1994jr}.

The primary backgrounds consist of atmospheric muons and neutrinos. The IceCube detector
itself can be used to veto muons inside of the volume of DeepCore, leaving only neutrino-induced showers to compete with.  For the spectrum of atmospheric neutrinos, we use the results of Ref.~\cite{Honda:2006qj}, which are in good agreement with the measurements of AMANDA~\cite{Collaboration:2009nf}.

The effective area of the detector for neutrinos can be defined as
\begin{equation}
A(E) \approx \rho_{\rm ice} N_A \sigma_{\nu N}(E)V(E),\label{eq:effarea}
\end{equation}
where $\rho_{\rm ice} =0.9 ~\mbox{g/cm}^3$, $N_A = 6.022 \times 10^{23}~\mbox{g}^{-1}$ (to convert grams to nucleons), $\sigma_{\nu N}(E)$ is the neutrino-nucleon 
cross-section~\cite{Gandhi:1998ri} and $V(E) \approx 0.04$ km$^3$ is the effective volume of the detector for a neutrino-induced shower of energy $E$~\cite{Resconi:2008fe}.

The directional capability of IceCube for a neutrino-induced shower above 1 TeV is expected to be on the 
order of $50^\circ$. We conservatively consider the signal and background over a solid angle corresponding to a full half of the sky ($2 \pi$ sr), acknowledging that our results would be strengthened if better angular resolution could be obtained. Using an NFW profile, we integrate the DM distribution in the direction of the Galactic center over this solid angle. We take the energy resolution of the detector to be $\log({{\rm E_{max}} / {\rm E_{min}}}) \sim 0.3$  \cite{Resconi:2008fe}.



As a first case, we consider a WIMP which can potentially provide the rising positron fraction observed by PAMELA~\cite{Adriani:2008zr}, as well as the electron spectrum observed by FGST~\cite{Collaboration:2009zk}. This requires a very heavy WIMP ($m_{\chi} \gsim 1$ TeV) which annihilates preferentially to muons (as opposed to electrons or taus)~\cite{Bergstrom:2009fa}. In Table~I and Figure~I, we show the boost factors to the annihilation rate that would be required for such a WIMP to be discovered or excluded by IceCube/DeepCore and compare this to the boost factor that would be required to produce the PAMELA and FGST signals~\cite{Bergstrom:2009fa}. We find that any WIMP that is capable of generating the FGST and PAMELA signals will also be well within the reach of IceCube/Deepcore.  In Figure I, we only plot the results for DM annihilating to muons.

Next, we turn our attention to a DM particle capable of generating the positron excess observed by PAMELA, without requiring that it also produces the spectrum reported by FGST. The results are shown in Table II and Figure~I.  In this case, although the full range of annihilation channels and masses capable of providing the PAMELA signal cannot be tested by IceCube/DeepCore, a significant fraction of the models can be. In particular, a 500 GeV WIMP which annihilates largely to muons or taus will be near or within the $2\sigma$ reach of IceCube/DeepCore for the entire range of boost factors capable of producing the PAMELA signal (this range corresponds to uncertainties in the cosmic ray propagation model). If the WIMPs also have an annihilaion channel directly to neutrinos, the reach is further 
extended. Lighter WIMPs are more difficult for IceCube/DeepCore to constrain or detect.


It should be noted that our results could potentially be improved upon if the angular resolution of IceCube/DeepCore turns out to be considerably better than we have assumed here.  Whereas the backgrounds are distributed broadly over the entire solid angle considered, the signal is concentrated in the region around the Galactic Center, thus enabling much greater statistical power if the angular window were to be reduced. Furthermore, as the $\nu_{\mu}$ background is considerably larger than that from $\nu_e$'s, any discrimination between electromagnetic and hadronic showers could be used to reduce the backgrounds and improve the statistical reach of IceCube/Deepcore to the signal discussed here. Although the results presented here assume that such discrimination is not possible, we remain hopeful that this will improve in the future.

In summary, in this letter we have calculated the flux of neutrinos from dark matter annihilations in the halo of the Milky Way in scenarios in which such annihilations are responsible for the cosmic ray positron excess observed by PAMELA and/or the cosmic ray electron spectrum observed by the Fermi Gamma Ray Space Telescope (FGST). To generate the PAMELA or FGST signals, dark matter particles must annihilate at a very high rate relative to that predicted for a typical thermal relic, and must annihilate primary to leptons. We have found that the neutrinos produced through such dark matter annihilations are likely to produce a signal observable in the low threshold extension of IceCube, known as DeepCore. With this goal in mind, IceCube itself will be used to veto atmospheric muons, while DeepCore will detect and identify neutrino-induced showers.

We find that, in any scenario in which dark matter annihilations produce both the PAMELA and FGST signals, IceCube/DeepCore should be capable of detecting corresponding neutrinos with greater than 5$\sigma$ significance. In a scenario in which only the PAMELA excess results from dark matter annihilations, IceCube/DeepCore will be capable of placing stringent constraints, but will likely not be able to exclude the entire range of possible dark matter masses and annihilation modes. As a final note, IceCube/DeepCore could also effectively probe decaying DM scenarios, though we have deferred such a discussion for the sake of brevity.

\medskip

We would like to thank A. Aguirre, P. Gondolo, K. Hoffman, S. Profumo, F. Halzen, I. Mocioiu and 
especially Spencer Klein for useful discussions.
KF is supported by the US Department of Energy and MCTP via the Univ.\ of
Michigan (K.F.); DS is supported by NSF grant AST-0507117 and GAANN (D.S.); DH is supported by the US Department of Energy, including grant DE-FG02-95ER40896, and by NASA grant NAG5-10842. KF thanks the Physics Dept. at the Univ. of Texas for hospitality
and the National Science Foundation under Grant No. PHY-0455649 for support. KF thanks
the Miller Institute at UC Berkeley (where this research was initiated) for support during her stay as Visiting Miller Professor. MRB was supported by DoE DE-FG03-92-ER40701.

\end{document}